# Cell response in free-packed granular systems

*Ana F. Cunha[1], André F. V. Matias[2,3], Cristóvão S. Dias[2,3], Mariana B. Oliveira[1]\*, Nuno A. M. Araújo[2,3]\*, João F. Mano[1]\**

[1]Department of Chemistry, CICECO – Aveiro Institute of Materials, University of Aveiro, 3810-193 Aveiro, Portugal;

[2]Centro de Física Teórica e Computacional, Faculdade de Ciências, Universidade de Lisboa, 1749-016 Lisboa, Portugal;

[3]Departamento de Física, Faculdade de Ciências, Universidade de Lisboa, 1749-016 Lisboa, Portugal

ABSTRACT

The study of the interactions of living adherent cells with mechanically stable (visco)elastic materials enables understanding and exploiting physiological phenomena mediated by cell-extracellular communication. Insight on the interaction of cells and surrounding objects with different stability patterns upon cell contact might unveil biological responses to engineer innovative applications. Here, we hypothesize that the efficiency of cell attachment, spreading and movement across a free-packed granular bed of microparticles depend on microparticle diameter, raising the possibility of a necessary minimum traction force for the reinforcement of cell-particle bonds, and long-term cell adhesion. The results suggest that microparticles with 14-20 µm are prone to cell-mediated mobility, holding the potential of inducing early cell detachment, while objects with diameters from 38-85 µm enable long-lasting cell adhesion and proliferation. An in-silico hybrid particle-based model that addresses time-dependent biological mechanisms of cell adhesion is proposed, providing inspiration for engineering platforms to address healthcare-related challenges.

**Introduction**

Granular packing spontaneously occurs in nature (e.g. sand, snow), and granular systems have been developed to ease packing in several fields of application including agriculture, food industry, construction and pharmacology[1,2]. Spherical particulate systems have gained interest due to their self-arrangement ability to attain a static equilibrium, whilst behaving as a fluid upon the application of external stimuli[1,2].

Individual particles for bioengineering have been mostly used as carriers for drug delivery[3] or as building blocks for tissue engineering to support cell adhesion and growth[4]. Several studies have focused on nanoparticle internalization[5,67], since nanosized particles can be uptaken by almost all cell types[8]. Particles can be endocyted[5,6,9,10] through two processes: (i) pinocytosis (15 to 200 nm) and (ii) phagocytosis (250 nm up to 19 µm), the latter mostly associated to cells of the immune system[11,12]. Although particle uptake efficiency varies with shape and size[9], an optimal phagocytosis interestingly corresponds to the size of most bacteria[13]. Even though mesenchymal stem cells (MSC) do not usually enroll in endocytosis of larger micron-sized particles[14], they were reported to sustain long-term internalization of 1 µm polystyrene and drug-loaded PLGA particles[15–17].

On a larger scale, microparticles have been mostly explored as cue-providing surfaces for cell support[3], with applications as supports for large-scale cell expansion, as well as cell carriers to integrate injectable scaffolds[18–22]. Multiple microparticles' properties have been explored, including biophysical aspects such as topography, geometry, stiffness, porosity, and area/volume ratio, as well as biochemical features capable of directing cell response[3]. Microparticles within size ranges of c.a. 50 µm have been confined in hollow capsules to provide cell-adhesive substrates for the formation of hybrid cell-particle clusters targeting tissue regeneration[23,24]. Moreover, micro-objects with different shapes (cubes, parallelepiped and crosses), with 40 µm lateral dimensions, were used to fabricate cell-object aggregates that can self-organize into geometrically controlled macrostructures[25,26]. Hydrogel/microgel jammed aggregates have been explored as injectable units for 3D printing, as well as *in situ*-forming hydrogels for drug delivery and micro/macro-tissue aggregation[20,27–30]. Particle size in jammed hydrogel microparticles may be tuned and, therefore, the porosity of the final assembled structure and cell response may also be tailored[21,31]. Mostly cell-free, but also cell-laden microgels, have been assembled to generate porous macrostructures that enable cell infiltration[21,31–33].

Although the behavior of several cell types in contact with individual or jammed/aggregated microparticles has been previously studied, individual cell response to particles organized in a mobile free-packed manner has not been explored so far. Recently, the interaction between free particles and 3D microtissues was suggested as a model system to understand the dynamics of cell invasiveness and internalization[34,35]. Here, we focus on the contact between individual cells and free objects of roughly the same size (Figure 1a). We hypothesize that in vitro free-packed microparticles of this size range may provide a quasi-3D platform where cells are allowed to interact with elastic objects, although in quite a fluidized environment, where cells are expected to pull the microparticles and move within this medium. Interestingly, although cell interactions with particles amenable to be internalized by cells (mostly < 5 µm) and particles capable of withstanding cell attachment (> 40 µm) are explored in a high number of studies, cell response in contact with particles in between those values still lacks characterization. Addressing this behavior may be of utmost importance to establish a minimum size of microparticles that could support cell attachment. This concept could also be useful in specifying injectable or bioprintable systems that would allow for a less invasive delivery of cargo-containing microparticles or to more precise 3D printed constructs, minimizing clog occurrence in needles and ejecting systems. Additionally, insight on interactions between inert and living matter may pave the way to mimic and understand phenomena including tissue intravasation, penetration and growth in healthy and disease settings.

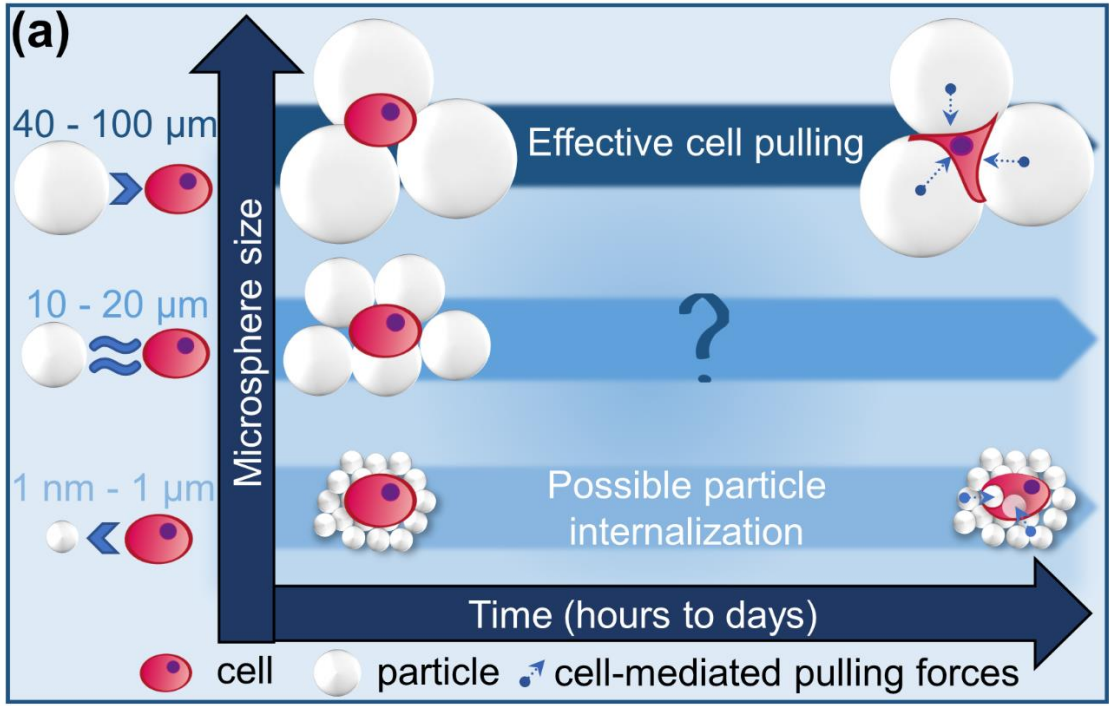
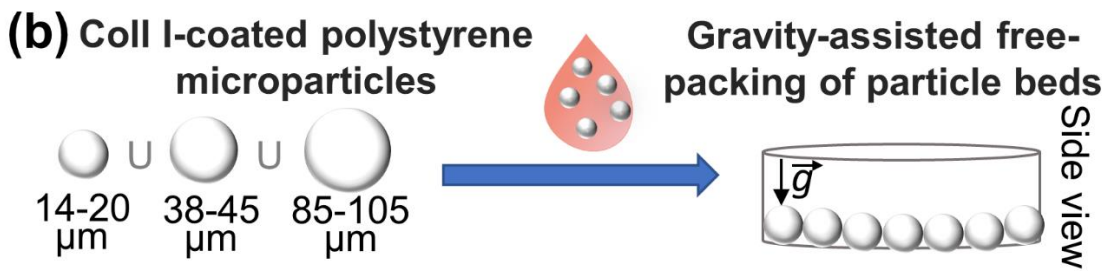
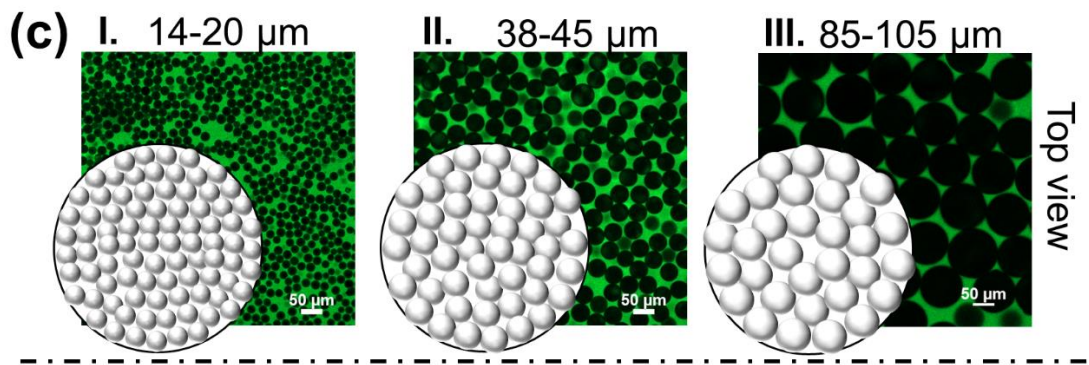
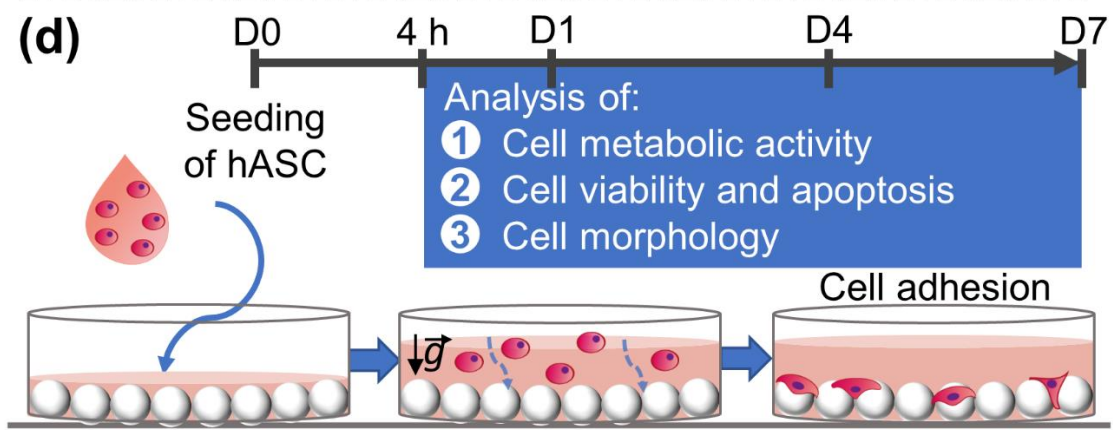

**Figure 1.** Schematic review of the differential cellular morphology and behavior on free-packed microsphere beds overtime, depending on microsphere sizes. (a) Nanosized particles can be internalized by almost all cell types whilst only phagocytic cells are able to endocytose microparticles[5]. Larger microparticles are sensed as surfaces for potential cell support or stimuli, therefore their biophysical and chemical properties can be tuned to modulate cell fate[3]. In this study, we address adherent cell response while in contact with free-packed beds of spheric microparticles, including the ones within the often overlooked diameter of 10 to 20 µm. (b) Three size ranged microparticles were coated with collagen I before free-packing assembly to obtain particle beds visible in confocal microscopy images shown in panel c, where only the top layer was focused. (d) hASCs were seeded on top of the bead beds and their response was evaluated for 7 days.

## Results

**Design of a cell-free packed microparticles model and exploitation of the hypothesis.** We suggest a model to study the time- and cell motility-dependent response of a hybrid granular system composed of free-packed microparticles and living individual cells. The response of human mesenchymal stem cells derived from the adipose tissue (hASC) was characterized while in contact with free-packed mono- or multilayers made of spheric microparticles with three well-defined diameter ranges: 14-20 µm, 38-45 µm, and 85-105 µm (Figure 1c, d). hASCs were selected as a source of clinically relevant primary cells due to their differentiation potential into different tissue cell types, as well as paracrine signaling and immunomodulatory properties widely explored for regenerative therapies[36–40]. Polystyrene (PS) is well-established as a non-biodegradable standard polymeric material for in vitro cell expansion[41]. PS 1.07 g cm$^{-3}$ density is close to the one reported for cells and, importantly, also similar to polymers commonly used as microparticulate constituents for cell expansion and tissue engineering approaches (e.g., PCL - 1.142 g cm$^{-3}$)[42]. Collagen type I was adsorbed onto microspheres' surface to provide cells with native ECM-mimetic domains, known to be bound by cells via integrin receptors[43,44] (Figure 1b).

We raise the hypothesis of a limitative traction force necessary for cell-mediated particle mobility, controlled by microparticles' size and density, potentially affecting the cells' ability to adhere and remain attached to spheres overtime (Figure 1a). Cell adhesion to the ECM or to biomaterials is initially mediated by integrins[45]. Such bonds, localized at the cell membrane, lead to the subsequent generation of intracellular forces that regulate adhesion, spreading and maturation, triggering downstream signaling cascades[45–47]. Talin is a key cytoplasmic protein available in its autoinhibited conformation[48,49]. When unfolded, talin reinforces the nascent integrin-ECM adhesions by linking actin cytoskeleton and integrins via vinculin recruitment[47,50–52]. This will stabilize actomyosin fibers which, in turn, exert force essential for the maturation into a focal adhesion[46,48,53]. Without talin reinforcement, the adhesion disassembles, ultimately causing cell detachment[54,55]. We expect that the forces established between a cell and sufficiently small-sized microparticles may enable substrate mobility, culminating in the inability of cells to successfully form mature bonds, possibly leading to cell death through anoikis (an orchestrated cell death due to the lack of a supportive ECM contact[55]).

**Characterization of hASCs cultured on a hybrid-granular system.** Distinct cell morphologies after one day of culture could be detected on the three size ranges of microspheres (Figure 2a and 3a). with increasingly stretched cytoskeleton observed with increasing microparticles' diameter. On 14 µm microparticles, cells did not develop stretched actin fibers, as opposed to the other size-ranged spheres (Figure 3a). Initial cell adhesion, as well as apoptosis markers, were detected in all experimental conditions at initial stages of incubation (day 1, Figure 2a). After 7 days of culture in a substrate of 14 µm spheres, cells not only failed to strongly adhere to the particles (day 1, Figure 3a), but also the detection of a viability marker (calcein) was decreased, along with the increase on apoptotic staining conferred by a phosphatidylserine marker (Figure 2a). This trend suggests that smaller spheres may also trigger an early apoptotic state on cells (Figure 2a). On the other hand, medium-term adhesion (up to 7 days) was observed for both 38 and 85 µm systems, which most probably correlates with the effective cell anchoring onto the substratum of spheres.

A semi-quantitative analysis of microscopy images stained for viable, early apoptotic and dead cells confirmed that while on initial adhesion time points, 4 hours and 1 days, all conditions showed a high proportion of living cells. After four days of incubation, smaller microspheres presented a lower ratio of living cells, along with a higher number of apoptotic cells (Figure 2b). This tendency was maintained for 7 days of cell culture, while in counterpart larger microparticles an evident increase on apoptotic cells could not be observed overtime.

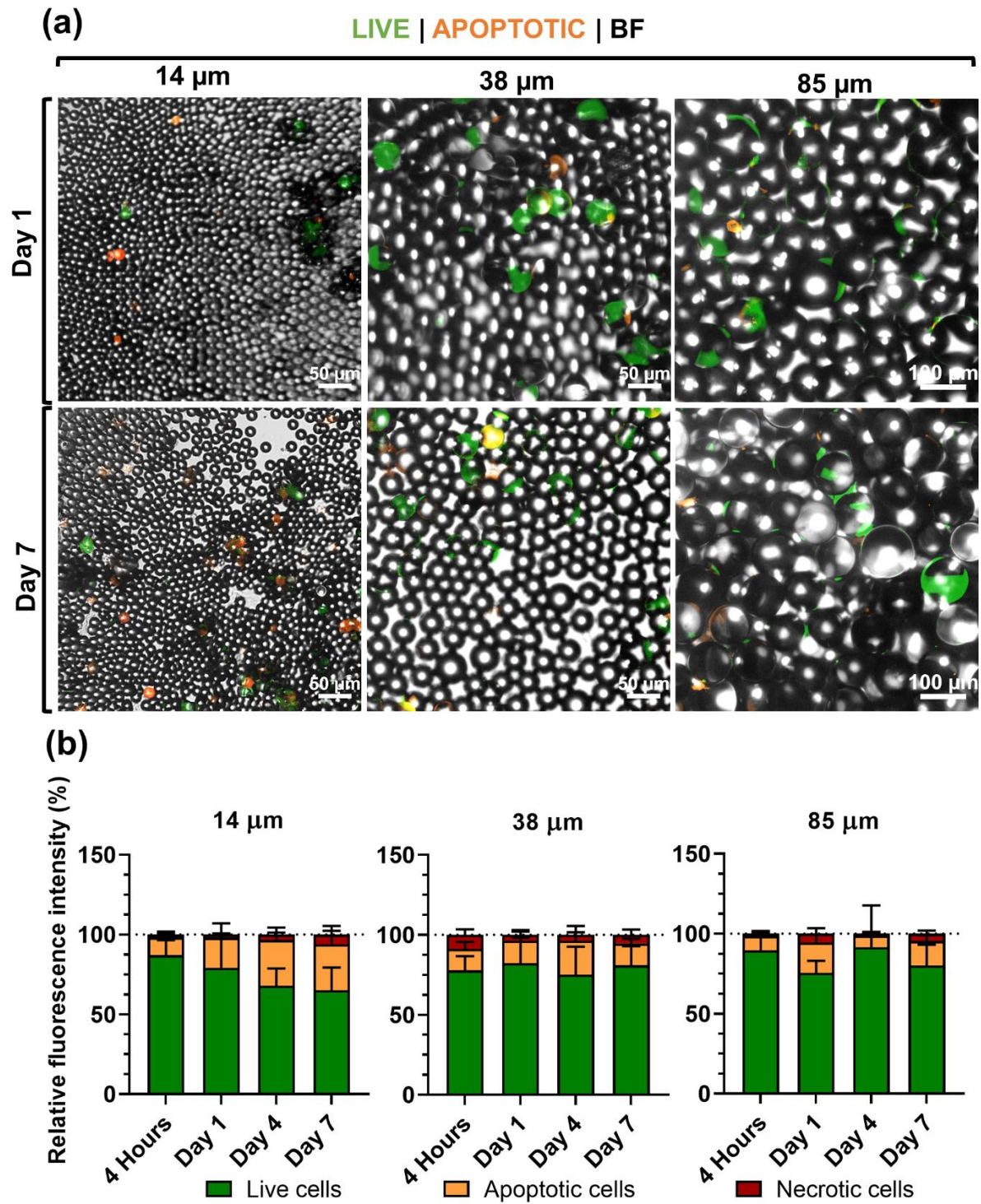

**Figure 2.** (a) Fluorescence microscopy images of live and apoptotic hASCs after 1 and 7 days of culture on top of particle beds. The first column displays the detection of an early apoptotic marker (orange), while viable cells are shown in green. (b) Fluorescence microscopy images of live,

apoptotic and necrotic stained cells analyzed for the relative fluorescent area of each staining (%); n = 4 independent experiments, 2 images/condition/timepoint. Data presented as means±s.e.m.

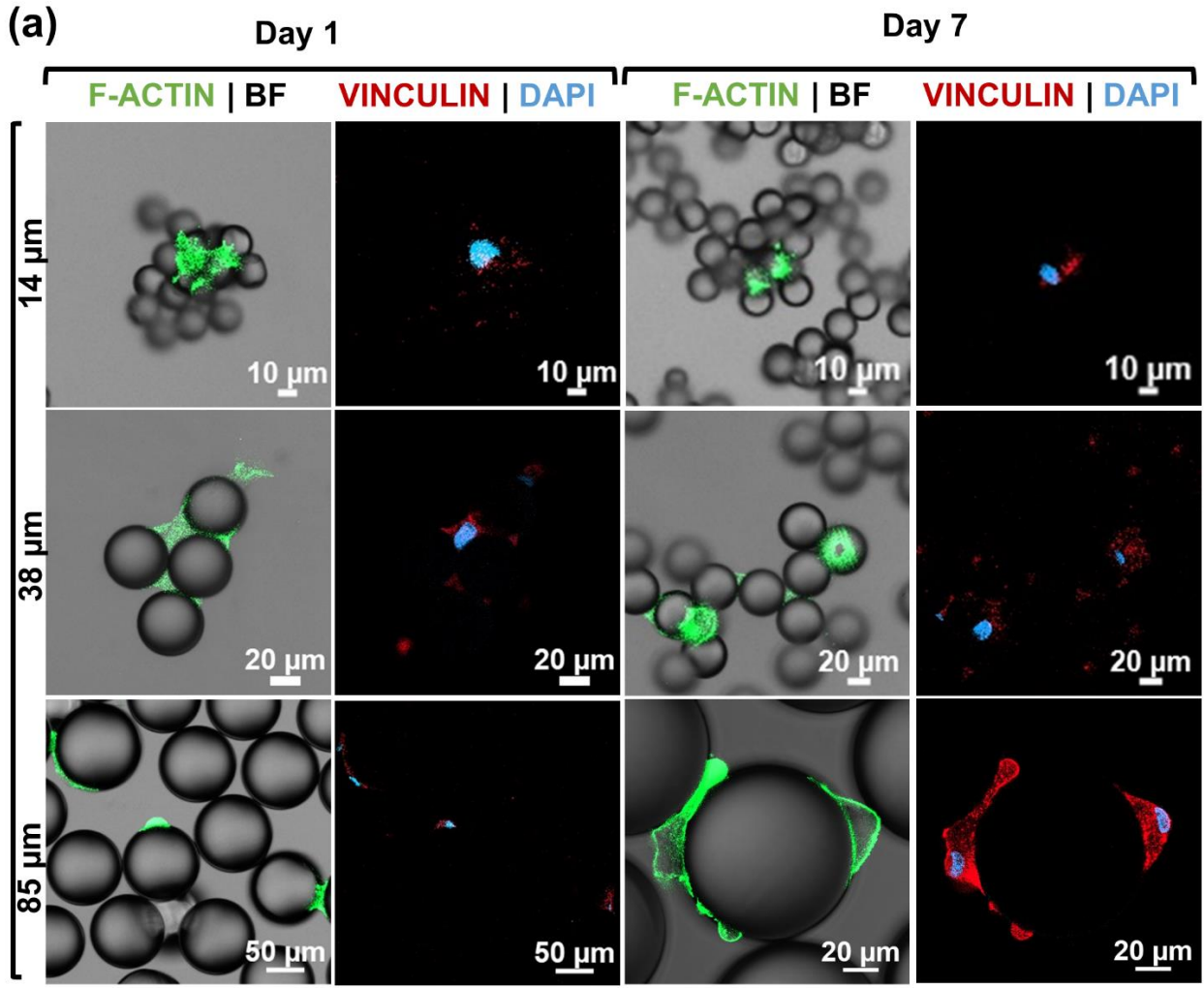

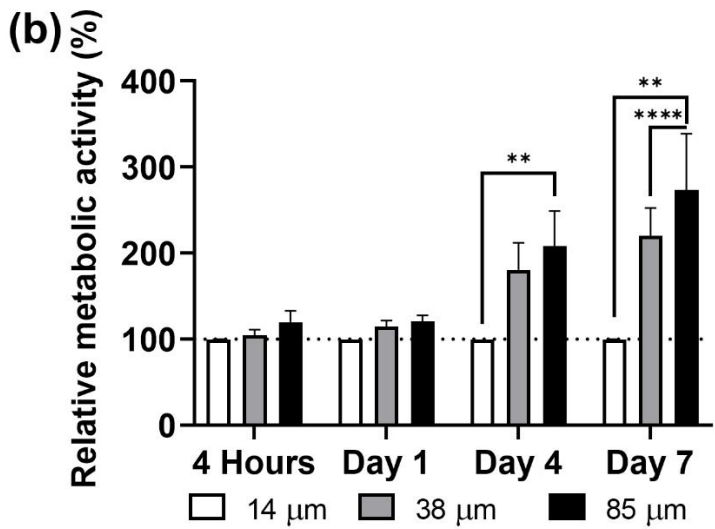

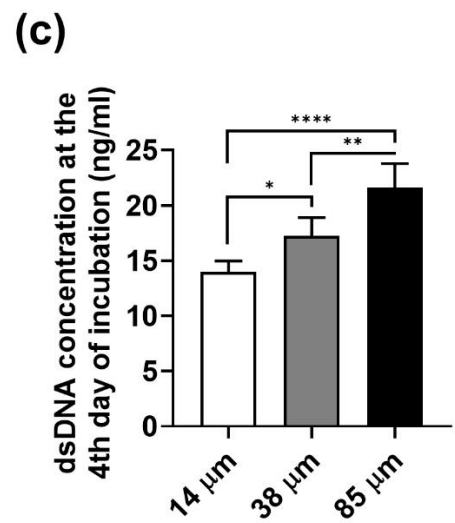

**Figure 3.** Free-packed microspheres of different size ranges (14-20 µm, 38-45 µm, 85-105 µm) up to 7 days in contact with hASCs provide different surface stability due to cell's pulling forces, compromising cell permanence, long-term anchoring and proliferation. (a) The first two columns were acquired at the first day of incubation and the 3$^{rd}$ and 4$^{th}$ column are after a week in culture. The first column displays the staining of F-actin and the second column shows detection of vinculin and cell nuclei. (b) Relative metabolic activity assessment of the spherical culture conditions for 7 days, compared to the first timepoint. N = 6 independent experiments (5 replicate wells/experiment), Two-way ANOVA with multiple comparisons for each timepoint. Data presented as means±s.e.m. (c) Quantification of the dsDNA content (ng/ml) of hASCs after four days in culture with different sized collagen-coated microparticles. One-way ANOVA with Tukey multiple comparison's test, n = 5 replicate wells. Data presented as means±s.d. In b and c, statistical significance was considered when $*P < 0.05$, $**P < 0.01$, $***P < 0.001$, $****P < 0.0001$.

The analysis of overall fluorescence associated to living and apoptotic cells showed a clear tendency for a decrease of the total number of cells adhered to 14 µm microparticles after four days of culture (Figure 2a), strongly suggesting that apoptotic cells present at day 1 may have undergone detachment from particles. Quantification of total dsDNA, directly proportional to cell number present on particles, corroborated that, indeed, at day four of cell culture 14 µm microparticles showed a statistically significant lower number of cells when compared to the systems comprising larger beads (Figure 3c). Cell metabolic activity analysis corroborated the tendencies observed for the presence of viable cells on microparticles and confirmed by dsDNA quantification. Four hours after cell seeding, the metabolic activity of all experimental conditions was similar (Figure 2b), suggesting that initial cell adhesion occurred to microspheres of all sizes to a similar extent. This result was kept for 24 hours of cell culture. A significant increase on the metabolic activity of cells cultured on spheres with average diameter sizes of both 38 µm and 85 µm was then observed, at days 7 and 14, compared to control 14 µm microparticles (Figure 2b).

Altogether, data retrieved from fluorescence microscopy and metabolic activity assessment suggest that nascent adhesions formed at the leading edge of cells' cytoskeleton allows their adhesion to microparticles with diameter in the range of 14-20 µm. However, because these particles are free-packed and have freedom of movement, they fail to provide the stability and space necessary for the stretching and polymerization of actomyosin fibers, essential for the maturation of the initial adhesion. Indeed, this range of particles' dimensions and weight seemed to correlate with a failure to reaffirm cell-particle adhesion forces resulting in cell detachment. Therefore, this experimental condition could have promoted an automated cell detachment and possible subsequent death due to the lack of contact with a stable matrix/surface after cells entered in an apoptotic state. In fact, a higher prevalence of apoptosis was detected in the 14 μm condition, considering the ratio of live/necrotic/apoptotic cells present on the particles (Figure 2b). When considering microparticles/objects with dimensions higher than 38 μm, hASCs are probably capable of forming transient adhesions at the extremities of the cytoplasm, that upon talin reinforcement develop into focal adhesion complexes allowing the meanwhile stretched cell to be fully anchored to the surface. These results further validate the usefulness of objects ranging from 40 to 100 μm as carriers to sustain cell adhesion and proliferation in tissue engineering studies, as reported in the literature[23–25].

To better understand and prove the dependency of cell response on two time-dependent and subsequent phenomena, consisting of (i) the establishment of initial cell adhesion to particles and (ii) the subsequent reinforcement of such adhesion to enable medium- or long-term cell adhesion, experimental and numerical approaches were explored. To assess the role of actin stretching as a necessary step for the establishment of long-lasting cell adhesion, actin contractility was inhibited experimentally on the free-packed system comprising larger particles. Additionally, to prove the role of small particles' mobility on cell detachment, a sintering model was developed both experimentally and on a numerical model designed to impair particle mobility.

**Inhibition of cell contractility:**

**(i) Addressing the role of adhesion maturation.** ROCK is a kinase responsible for stress fiber contraction as well as F-actin stabilization through myosin binding[47], and can be inhibited by Y-27632 compound to disassemble actin fibers[56]. For 85 μm particle beds, a significant decrease on cell metabolic activity occurred after four days of cell culture, which was not observed in the untreated condition (Figure 4b). Fluorescence staining of F-actin showed thinner filamentous actin cytoskeleton and nascent adhesions that were unable to mature due to the presence of this compound (Figure 4a). Thus, we conclude that Y-27632 treatment weakened actin fibers force generation leading to cell detachment in particles that, otherwise, behaved as ideal substrates for cell adhesion, stretching and proliferation (Figure 3a). This trend was corroborated using a more downstream inhibitor of myosin II – blebbistatin[57] that halts actomyosin assembly (Figure 4a and b). The presence of this inhibitor causes a distinct cell organization with a cytoskeleton arranged in thin filaments surrounding cell nucleus and organelles possibly, as cells appear weakly adhered to the particles.

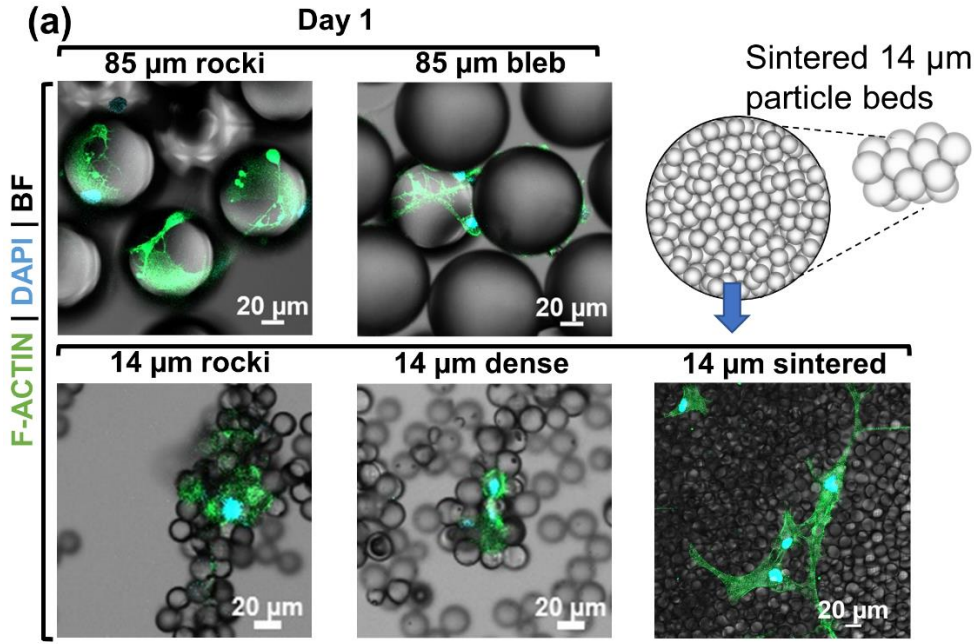

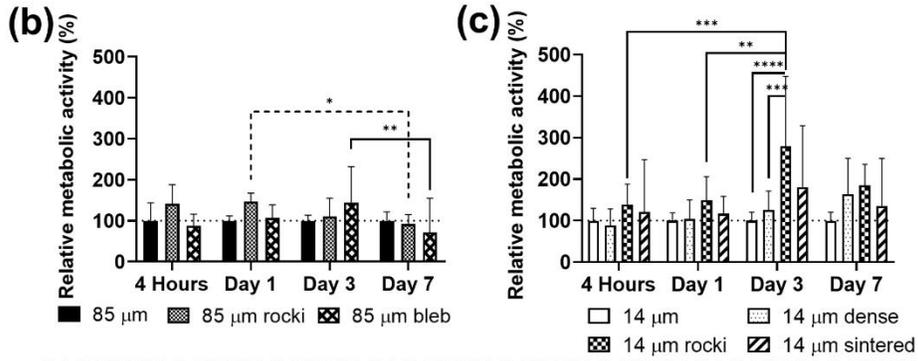

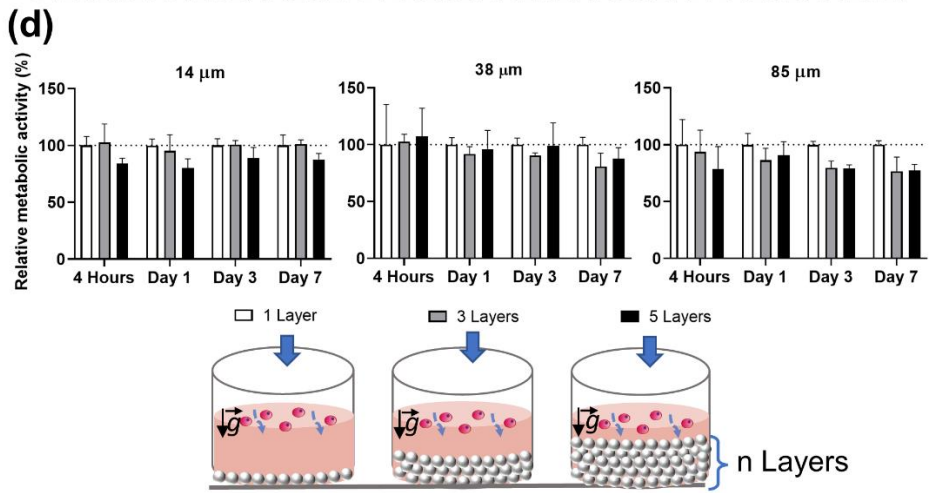

**Figure 4.** (a) Staining of actin filaments cell nuclei, after 1 day in culture with different size-ranged particles to assess cell response to the inhibition of cell contractility and particle mobility. (b) Metabolic activity assessment of hASCs through 7 days in culture with 85 µm collagen-coated microparticles with and without ROCK and blebbistatin inhibitor treatment (Y27632, 10 µM, bleb 25 µM) displayed. Two-way ANOVA with multiple comparisons was performed on n = 15 replicates. (c) Evaluation of the metabolic activity of hASCs cultured in controlled 14 µm particle beds: free, denser, rock inhibitor treatment and sintered (n = 15 replicate wells, obtained in 3 independent experiments, Two-way ANOVA with multiple comparisons). Inset showing a schematic representation of 14 µm sintered beds. (d) Effect of number of layers in free-packed particle beads on cell metabolic activity after a week in culture. Statistical significance was considered when *$P < 0.05$, **$P < 0.01$, ***$P < 0.001$, ****$P < 0.0001$. All data presented as mean+s.d.

**(ii) Decreasing the tendency for the formation of actin fibrils**. We applied an inhibitor of ROCK for the 14 µm condition, leading to a decrease in cell contractility, to modulate cells' ability to pull spheres (Figure 4b). A significant increase in cell metabolic activity was detected for 14 µm particles in the presence of the ROCK inhibitor for 3 days of culture, with a subsequent decrease at 7 days of cell culture. This suggests that a lower contractility of cells may correlate with their inability to effectively pull the particles, which has impaired their movement, enabling cells to remain attached to particles during the first 3 days of contact. In fact, the inhibition of the Rho-ROCK pathway is regularly used to improve cell viability of embryonic stem cells during freeze-thawing cycles[58,59]. However, MSCs have been reported to depend on the effective establishment of contractile forces to survive on the long term[60], which may explain the decrease observed in their metabolic activity at day 7.

**Experimental controls to address the role of particles' mobility:**

**(i) Microparticle sintering.** To assess if the inability to reinforce and mature bonds in free small particles can be ascribed to particle mobility mediated by cellular forces, a system comprising sintered particles was applied to reduce particle mobility. The sintering of 14 μm spheres promoted an increase in the metabolic activity at day 1, relative to the initial timepoint, as opposed to the experimental condition with mobile microspheres (Figure 4c). This finding reinforces the critical role of cellular forces in mediating particle-cell adhesion.

**(ii) Use of denser microparticles.** Collagen-coated soda lime glass microparticles with average diameter of 14 μm (2.5 g/cm$^3$ density) were used as a denser material to be directly compared to PS. Although collagen adsorption to glass particles may have occurred at a distinct extent than to PS particles, this condition could be useful to extrapolate about the ability of hASCs to pull microparticles with higher density, still in the range of biomaterials widely used for cell culture and expansion. We observed that, while for initial timepoints cells adhered to glass microparticles in comparable extents to the lower density PS particles, at day 7 a higher number of cells remained in glass denser particles, when compared to the polystyrene ones, therefore reinforcing the susceptibility of small particles to cell-imposed movement as a relevant mechanism explaining our results (Figure 4a and c).

**Effect of number of layers in free-packed systems.** Granular systems with 1, 3 and 5 layers prepared in flat bottom plates were compared. Although a higher number of layers was expected to impart higher stiffness to the overall system, culminating in lower particle mobility, a tendency for systematic lower metabolic activity was detected in all conditions for 3 and 5 layers of microparticles, when compared to monolayers (Figure 4d), with statistical significance for 85 μm microparticle systems. We hypothesize that the larger interstitial spaces (in between particles) in systems composed of larger microparticles may benefit the direct passage or migration of cells to bottom layers. In those inner regions, oxygen levels and medium renewal in static conditions used for culture may be lower, which may deaccelerate cell proliferation. Of note, this effect was mostly noted as a robust trend for 3 and 7 days of culture.

**Simulation model.** Recently, computational analysis has gained momentum as a complementary tool to understand and characterize cell adhesion processes[61,62] and their role in morphogenesis[63,64]. To shed light on the role of the cell/microparticles interaction we performed particle-based simulations. We propose a discrete model of a cell that can adapt to the bed of microparticles, adhering to the microparticles and spreading. We consider a simple mechanism of anchoring of protrusions, which allows us to determine in a quantitative way the impact of the size of the microparticles on the cell spreading.

We consider cells on a granular bed of spheres of mean diameter $d$ and with 5% size dispersion (Figure 5a). To obtain the configuration of the granular bed, we performed discrete element simulations to let the microparticles fall freely on a surface, yielding inelastic collisions and thus minimizing their kinetic and potential energies until they stop (see details on the Materials and methods section). Cells are represented by a set of spheres (elements) connected by viscoelastic springs (structural springs). Initially, each cell consists of seven elements, organized in a hexagonal configuration and the diameter of cell $L_i$ corresponds to the diameter of the smallest circle that contains all particles (see Figure 5b). For simplicity, we consider that the density of each element in the cell is the same as the one of each microparticle[42,65].

We model the cell spreading through cycles of stretching and contraction[66]. During stretching, a new layer of elements is added to the perimeter of the cell, representing the branching-out of protrusions. Per element of the periphery of the cell, two new elements are added in a new layer, and we allow enough time for the new layer to relax (net force zero) before contraction (see Movie S1, Supplementary Information). The contraction occurs by decreasing the natural length of the structural springs down to zero, representing the force generated by the actomyosin fibers. The length of the structural springs decreases at a steady rate. This cycle is repeated up to three layers of elements corresponding to a maximum cell diameter of $L_f$. In experimental units, we estimate the initial cell diameter to be 32 μm, this means that each cell can spread up to 64 μm.

We consider three binding/unbinding mechanisms in the process of cell adhesion (Figure 5c): initial cell adhesion; cell detachment; adhesion reinforcement. Cell adhesion is effective by the engagement of integrin to the ECM[45,66]. We consider that this process is instantaneous. Cell detachment is triggered by the disengagement of integrins, which occurs in the time scale of four seconds[51,54,55]. Assuming that a cell contracts at a constant velocity of 110 nm/s (as proposed in Ref.[51]), the cell can contract up to 4% of its initial diameter within the lifespan of the integrin bonds. Cell detachment is prevented if the adhesion is reinforced within the lifespan of the integrin bonds. The reinforcement occurs due to the unfolding of talin and subsequent reinforcement with vinculin[47,50–52]. The time required for this unfolding depends strongly on the force transmitted through the integrin bonds, *i.e.* the cell/microparticle interaction, and above a certain threshold, for the value of the force, it drops significantly. We consider that the process is instantaneous if the value of the force goes above a given threshold[51]. We set the threshold to be the maximum force a cell can exert (200 pN for the cell type chosen for the experiments[67]). We assume that this maximum force remains constant as the cell spreads[68].

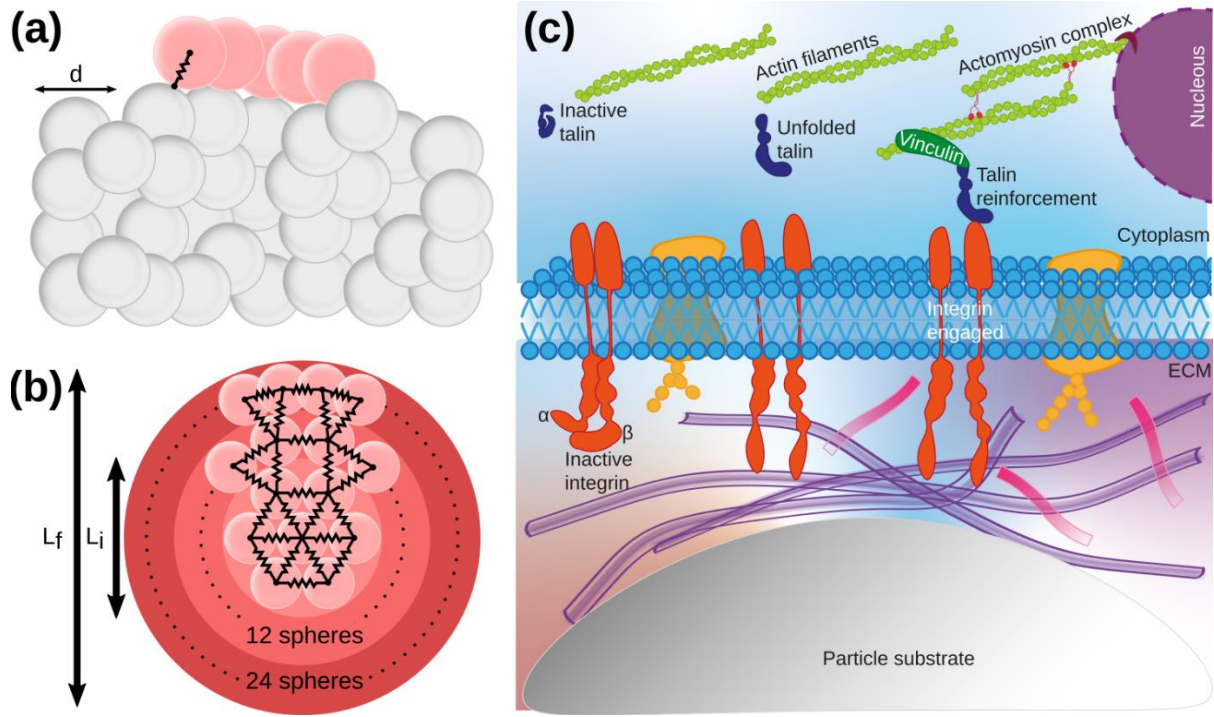

**Figure 5.** Schematic representation of the numerical simulations and the relevant biological mechanisms. (a) The cell with two layers of elements, in pink, falls freely on top of the bed. Upon contact, connections are formed between a microparticle and the cell, represented by the black spring. (b) Structurally, the cell is composed of a network of viscoelastic springs (structural springs) that connect the elements inside the cell. Initially the cell starts with one layer (7 elements) and diameter $L_i$ and, if bonded to the microparticles, spreads the subsequent shells up to a diameter of $L_f$. Cell contraction is achieved by reducing the natural size of the structural springs. (c) Schematic detail of the relevant biological mechanisms considered in the simulations. After contact with the microparticle the integrin engages, and it remains engaged for 2-4 s. During the engagement period, it occurs the talin unfolding and reinforcement, the latter is successful only if the force threshold is met otherwise the engagement period is surpassed leading to cell detachment.

The initial cell adhesion occurs whenever an element contacts with a microparticle. The resultant integrin binding is modeled as a viscoelastic spring (binding spring). The elastic constant of the binding spring depends on the maximum cell force and the maximum spring displacement (see details in the Methods section). To ensure that, when the microparticles are not moving, the cell reaches the maximum force before detachment occurs, we set the maximum displacement of the binding spring to a value lower than the unbinding threshold.

The mechanism of adhesion reinforcement is triggered if the force, exerted by the binding spring, exceeds the threshold. In that case, we assume that the formation of actomyosin fibers is successful and fix the position of the bonded microparticle and the element. Cell detachment occurs if a particular element contracts a distance larger than the threshold of 4%, measured from the center element. This causes the unbinding of the microparticle-element pair and the cell stops spreading in that direction.

The cell fate was quantified numerically by the ratio between the final cell volume and the estimated cell volume if there were no spring contraction, which is an upper bound for the cell volume. Thus, this ratio is minimal when the cell contracts completely and unity when it spreads to its maximum. We simulated the cell adhesion and spreading for 16 different initial positions over the bed and considered 10 different bed configurations, resulting in 160 samples per cell-microparticle aspect ratio $L_i/d$. For large ratios, the microparticles are light enough to be pulled by the cell and thus the cell contracts below the unbinding threshold and all cell-microparticle pairs unbind. This results in a low average volume ratio (see Figure 6 and Movie S1, Supplementary Information). For larger radii of the microparticle, the cell is no longer able to pull the microparticle as it contracts, and the cell-microparticle bonds are reinforced, which leads to an increase in the average volume ratio. One can then distinguish two different regions. For values of the cell to particle ratio below one (larger microparticles) the cell does not contract, which leads to high survivability. Above that value, the higher degree of contraction suggests a lower survival rate.

Numerical data predict that packing of cell-sized microparticles reverses the observed impairment effect on cellular stretching of their experimental counterpart substrate. In fact, cells maintain their initial size/volume in a sintered particle substrate scenario regardless of the granule size, indicating the maintenance of cell adhesion and possibly proliferation over the course of the simulation (Figure 6). Changing the microparticles densities impacted cell survivability. For microparticles of higher density, the ratio between maximum cell force and microparticle weight decreases, which makes it more difficult for a cell to move a microparticle and thus the cell adhesions are preserved.

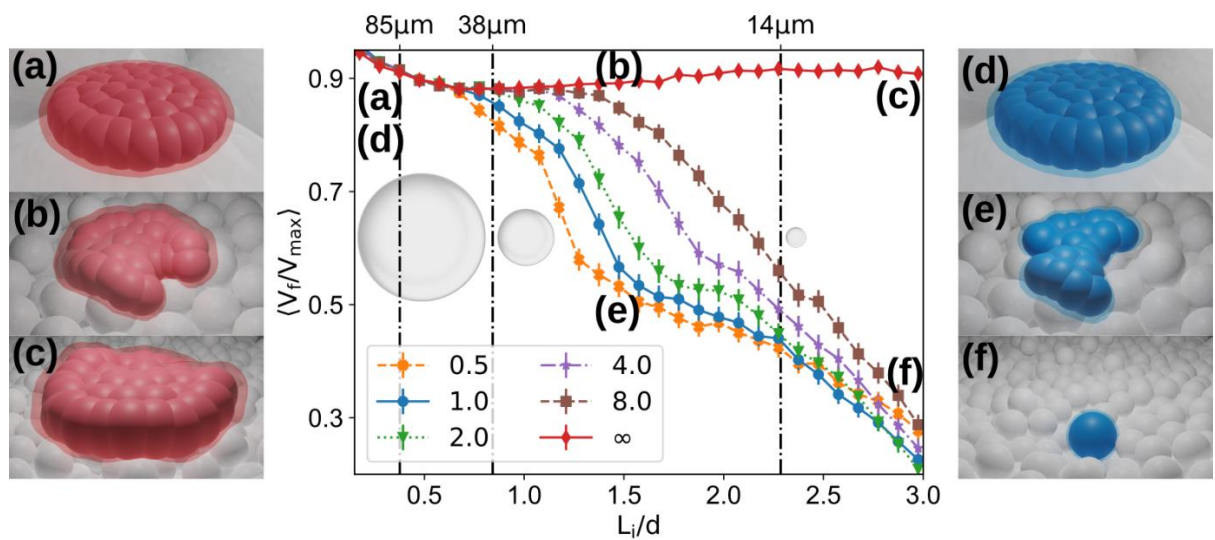

**Figure 6**. Numerical results of the cell contraction. In the middle, plot of the average ratio of final cell volume to maximum cell volume as a function of the cell-microparticle ratio. The different colors and markers correspond to different microparticle densities relative to polystyrene (blue circles). A relative density of 2 (green triangles) corresponds to the glass microparticles, and the ∞ (red diamonds) to the sintered microparticles. The error bars correspond to the standard error (n=160). The vertical lines correspond to an estimated conversion between numerical and experimental cell size, see top ticks. The left (right) snapshots represent the final cell for sintered (polystyrene) substrate microparticles for different cell-microparticle ratios, marked in the plot with the letters.

In silico simulations (Figure S7) also predicted negligible differences in cell permanence on the granular bed for systems composed of 3 or 5 layers, in agreement with the experimental data. For one layer, simulations predicted a slightly lower cell permanence for 85 μm microparticles, due to a lower interlocking, facilitating particle mobility. In fact, in vitro results showed that there were no detectable differences between using 1, 3 or 5 layers for the first timepoints post-seeding (4 hours and 24 hours), indicating that the numbers of cells adhered to the spheres is similar.

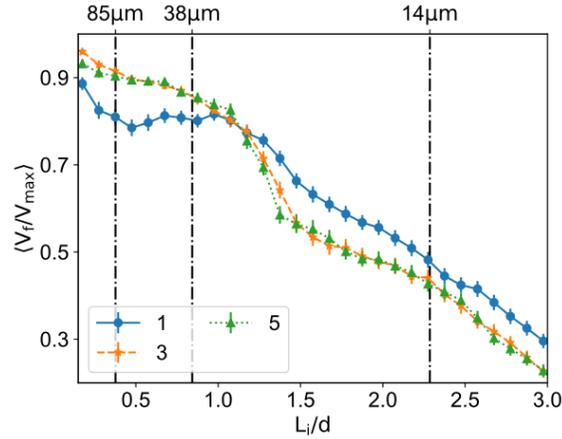

**Figure 7**. Numerical results with different number of microparticle layers. Plot of the average ratio of final cell volume to maximum cell volume as a function of the cell-microparticle ratio. The different colors and markers correspond to different number of microparticle layers. The error bars correspond to the standard error (n=160). The vertical lines correspond to an estimated conversion between numerical and experimental cell size, see top ticks.

**Conclusion**

Forces generated by adherent cells on surfaces are essential to understand complex biological processes including cell migration, developmental processes, tissue formation, and tissue integrity maintenance[69,70]. Granular systems are used for numerous applications being dynamic surfaces portraying both static and fluid-like environments. The degree of mobility of microparticles in the granular packs, dictated by intrinsic factors such as particle size and density, jamming efficiency, as well as exogenous applied forces (e.g., by adherent cells) may dictate the resemblance of these systems to softer or stiffer 3D substrates. In fact, in 3D experiments using hydrogels, low stiffness biomaterials correlated with the increased turnover of adhesion proteins due to decreased ability to form stable adhesions; in opposition, stiff 3D substrate enabled the formation of stable cell adhesion formation[60].

The design of a hybrid cell-granular system composed of cells (hASC) and free-packed collagen-coated polystyrene microparticles allowed a close view on the interface of cells and objects within specific diameter ranges (14-20 µm, 38-45 µm, 85-105 µm). Cell morphology and viability were assessed from early cell-material contact timepoints, up to seven days of culture. We observed that there is a scale of small microparticles that can neither be internalized or provide a stable surface for cell attachment. While large sized microparticles (85-105 µm) provided cellular anchoring and proliferation, 38-45 µm appears a promising size range for cell growth, despite rendering a cell response profile with higher variability. The reduction of cell-sized sphere substrate mobility through particle sintering warranted cell survival, while weakening cellular contractility with a ROCK inhibitor treatment on larger sized beads negatively impacted cell attachment to this otherwise favorable substrate.

The analysis of experimental and numerical data obtained by the developed modelling system corroborated that there is a competition between the force necessary for clutch reinforcement and the required force for particle mobility mediated by cells, that is time-dependent and determined by the object's biophysical characteristics. For medium-term cell adhesion, after initial integrin-mediated cell binding to materials, an effective strengthening of the cell-material bond seems to be imperative to enable effective cell attachment and proliferation. Otherwise, cytoskeletal disassembly and cell detachment take place. Following our premise, small-sized spheres are lighter and thus more easily pushed upwards by the cell rather than a particle-clutch reinforcement is to occur. Therefore, granular surfaces of small microparticles are more prone to convey cell detachment and impair cell proliferation.

Biological systems are complex non-equilibrium moieties and, therefore, very difficult to simulate. The model suggested here was designed considering time-sensitive mechanisms as parameters to explain the different cell response on different free-packed systems. Determining a minimum particle size for cell adhesion in hybrid-granular systems might be impactful in understanding and mimicking, for instance, tissue intravasation, and to design engineering systems with optimal injectability and printability. Reducing the size of microparticle for cell support also increases the overall surface area available for cell attachment; in fact, total surface area and the total volume (or mass) of particles scales with $r^{-1}$, where r is the radius of the particles. Also, particles systems within unexplored smaller non-internalizable size ranges may be interesting platforms to impair cell growth, tissue invasion and matrix deposition in pathological scenarios including cancer. Further assessment of the effect of different types of granular free-packed systems on stem cell phenotype (e.g., differentiation into different lineages, or maintenance of stemness) may also introduce the use of these system as technically simplified modulators of cell response.

**Materials and Methods**

**Microparticle surface treatment.** Commercially available translucent polystyrene microspheres with diameters of 14-20 μm, 38-45 μm, and 85-105 μm, and with density of 1.07g cm$^{-3}$ were purchased from Cospheric. UV-ozone surface treatment (PROCLEANER 220, BioForce Nanosciences) was performed for 5 min (with a shaking step in between) before the overnight coating of spheres with 10 μg cm$^{-2}$ collagen type I solution from rat tail in 20 mM of acetic acid (Sigma-Aldrich®), at 4° C.

**Generation of microparticle beds.** For metabolic activity assessments, protein-coated microparticles dispersed in cell culture media were added to 96-well plate V bottom and allowed to pack for 1 hour before cell seeding. Exceptionally, for the studies concerning the influence of the number of microparticle layers, free-packed beds were prepared in standard 96 well plates for suspension cell culture (6 mm diameter), dispensing a sufficient number of particles to render the intended number of layers (1, 3 or 5) in a flat substrate. For image analysis (widefield fluorescence and confocal microscopy), the free-packed layers were prepared using the same methodology in angiogenesis µ-slides (ibidi®, 81501).

**Cell culture.** Human mesenchymal stem cells derived from the adipose tissue (hASCs) were purchased from ATCC®. Cells were expanded in culture tissue flasks in alpha-Minimum Essential Medium (alpha-MEM) supplemented with 10% fetal bovine serum, 1% antibiotic-antimycotic (Gibco™) and sodium bicarbonate (Sigma-Aldrich®). Cells were used for experiments at 80% confluency between passages from 4-6. After washing with Dubelcco's phosphate buffer saline solution (Gibco™) and cell trypsinization, $10^3$ cells in 10 µl were gently seeded on top of a monolayer of coated microparticles using an untreated angiogenesis µ-slide (ibidi®, 81501). Low cell seeding was chosen to study cell-ECM interactions via integrins and minimize the probability of cell-cell contact. After 1 hour of incubation at 37° C in a humidified environment of 5% $CO_2$, 40 µL of cell media was added.

**Cell contractility inhibition experiments.** To inhibit/weaken actomyosin contractility and stabilization, the Y27632 compound (ATCC® ACS3030™) - an inhibitor of ROCK I and II - was supplemented in the cell culture medium of 85-105 µm particle monolayers, after 1 hour of incubation and throughout the whole experiment at a final concentration of 10 µM. Additionally, (-)-blebbistatin (Sellek Chemicals S7099) was used at 25 µM. Cell characterization was done at the same timepoints as the untreated condition.

**Microparticle sintering.** To reduce particle mobility mediated by cellular machinery, small particle (14-20 µm) packing was promoted while maintaining a spherical topography. Polystyrene spheres dispersed in 70% deionized water/ethanol mixed with tetrahydrofuran (Sigma -Aldrich®) in a 2:1 ratio was deposited in a volume corresponding to four particle layers (protocol based on Ref[71] with modifications to obtain a slight melting effect). Multi-layered bead packs were allowed to dry completely before cell culture assays, using the same workflow mentioned for free-packed particles.

**Fluorescence microscopy observation and image analysis.** To evaluate cell viability, Abcam®'s Apoptosis/Necrosis Detection Kit was chosen and used according to manufacturer specifications. The three colored staining allows the detection of apoptotic cells (phosphatidylserine marker), late apoptosis/necrosis (7-aminoactinomycin D that detects loss of membrane integrity) and live cells (calcein). After dye incubation for 30 min, images from fluorescence microscope Zeiss Axio Imager 2 were acquired with Zen Software 2.6. Afterwards, Fiji ImageJ software was used to obtain the fluorescence percentage area for each channel. Two pictures of distinct observation fields for each condition and timepoint were analyzed. For staining of filamentous actin, vinculin and cell nuclei of previously fixed samples (4% paraformaldehyde), after 10 min permeabilization with PBS + 0.1% Triton X and 1 hour blocking with 1% BSA + PBS + 1% Tween, Vinculin anti-human primary antibody (42H89L44, Invitrogen, 1:50) was added for overnight incubation at 4ºC. After three washing steps of 5 minutes in PBS, cells were staining with rabbit IgG AlexaFluor 647 (A-21245, Invitrogen, 1:150) for 1 hour at room temperature in the dark. Cells were incubated with Flash Phalloidin™ Green 488 (Biolegend, 1:20) for 20 min and counterstained with DAPI (diamidino-2-phenylindole, Thermo Fisher Scientific™ 1 mg mL-1, 1:1000) for 5 min. Fluorescently labeled samples were observed in Confocal Microscope ZEISS LSM 900 and acquired images were analyzed using ZEN 3.1 software.

**Cell metabolic activity assessment.** A colorimetric assay based on resazurin reduction using AlamarBlue Cell Viability Reagent (Invitrogen™) was performed to determine cell metabolic activity. In viable cells, a metabolic conversion of the resazurin compound to resofurin results in a detectable color shift. AlamarBlue solution was diluted in cell culture medium in a 1 to 10 ratio and incubated with the samples for 12 hours. Then, fluorescence intensity was measured by the microplate reader (Synergy™ HTX; excitation 560 nm, emission 590 nm) in a black 96-well plate clear flat bottom polystyrene microplate containing 100 µL of the incubated solution in triplicate. Reagent blank was subtracted from each sample for the extrapolation of the metabolic activity. After each timepoint, samples were washed with cell medium and maintained in culture at 37° C, 5% $CO_2$.

**dsDNA quantification.** Quant-iT™ PicoGreen® dsDNA (Invitrogen™) was chosen for its highly sensitivity for detecting double stranded DNA content. 100 µL of each sample was added in duplicate to a white 96-well plate opaque flat bottom polystyrene microplate. After adding 100 µL PicoGreen reagent to each well, a microplate reader allowed the fluorescence evaluation (excitation 480 nm, emission 520 nm).

**Statistical analysis.** Data analysis was performed with Microsoft Office Excel and statistical tests with GraphPad Prism 8. Statistical significances were considered when *p* value < 0.05.

**Discrete element method.** The size of microparticles is generated with a dispersion of 5% and they are randomly distributed (without overlapping) inside a three-dimensional box of size $L_x \times L_y \times L_z = (20d)^3$. On the $x$ and $y$ direction we impose periodic boundaries and there is a surface at the bottom ($z$ direction), such that $z > 1$ for all microparticles. We generate enough microparticles to have a bed with a height of three microparticles. To obtain a loose random packing, we use a discrete element method algorithm, where the microparticles are subject to a gravitational force along the vertical direction $z$ (Supplementary Video S2). The trajectory of each microparticle is obtained using the Gear's integration scheme of 4th order[72,73]. The collision between microparticles is mediated by a spring with a different elastic constant for microparticles approaching and moving away, as proposed by Walton and Braun[72,74,75]. The choice of the elastic constants allows the control of the coefficient of restitution. Our microparticles have the elastic constant $Y_u = 5 \times 10^4 d$ and $Y_l = 5 \times 10^3 d$ for when they are approaching and moving away, respectively. This results in a restitution coefficient of $\sqrt{Y_l/Y_u} \simeq 30\%$, allowing for the fast relaxation of the microparticles. Also, the high values of elastic constants ensures that the overlap between microparticles is minimal. The interaction with the surface is the same as microparticle-microparticle collision. We do not consider rotational degrees of freedom. The bed is relaxed when the average kinetic energy of the microparticles is smaller than $2 \times 10^{-3}$. The elements that compose the cell have the same collision properties of the microparticle of the bed. The simulation timestep is $5 \times 10^{-5}$ in dimensionless units.

**Viscoelastic springs.** The parameters of the viscoelastic structural springs were adjusted based on the weight and radius of the elements, to ensure that the cell wets the surface of a microparticle of the diameter of the final cell $L_f$. The elastic spring constant changes the flexibility of the cell. For a high elastic constant, the cell is rigid and does not adapt to the rugosity of the substrate, while for a very low elastic constant the cell is flexible, but its surface area is not well defined. The elastic constant is set to $6.25 L_i^2$, which allows for the cell to adapt to the surface of a microparticle of diameter $L_f$. We ran independent simulations with different elastic constants (ranging different orders of magnitude) and arrived at the same conclusions. The damping constant of the structural springs is set to $6.25 \times 10^{-2} L_i^2$, which is strong enough to guarantee that the vibrations are in the overdamped regime[76]. We set the damping constant such that the vibrations are negligible in the timescale of cell adhesion.

For the springs that connect the cells to the microparticles, the elastic constant is the maximum weight an element can lift divided by $2.5 \times 10^{-5} L_i$ of the element radius, meaning that the maximum force is achieved when the spring stretches $0.0025\%$ of $L_i$, far less than 4%, the threshold for element detachment. The damping constant is the maximum weight an element can lift divided by $1.25 \times 10^{-2} L_i$.

**Numerical cell spreading.** To ensure the right placement of the new layers of elements, we compute the position of all elements from the beginning of the simulation. Initially, only the first layer of elements is active, meaning that all elements on the second and third layer have their mass reduced by a factor of $10^3$ so that their impact on the bed configuration is minimal. The viscoelastic properties are rescaled accordingly. These elements have a fictitious binding to the bed (for numerical stability). Before each cycle of cell contraction and spreading, we recalculate the binding of all elements to the microparticles. Before each cycle of spreading the masses of the elements of the new layer are set to the same value has the active elements.

**Maximum cell force.** The maximum force of the cell was determined from the experimental values of the adhesion force[67] and the density of polystyrene microparticles, 1.07 g cm$^{-3}$. The polystyrene microparticles are in an aqueous solution, which we consider the density of water, 1.00 g cm$^{-3}$. This means that on the ideal case of a cell concentrating all its force on a particle, exerting a force of 200 pN can lift a mass of 1.96 µg which corresponds to a microparticle diameter of 82 µm. Considering the cell size reported by Zheng *et al.* of 32 µm, we define the maximum force as the necessary to lift a microparticles whose diameter is 82/32 times the cell diameter[67].

ASSOCIATED CONTENT

**Supporting Information**

The following files are available free of charge at Additional graphs regarding semi-quantification analysis of fluorescence microscopy images and videos of the simulation model developed to illustrate cell spreading in different particle beds.

Influence of particle bed conditions on cell spreading: particle beds with different cell-particle aspect ratios in fixed (**a-c**) and free-packing conditions (**d-f**) (MP4).

Microparticle substrate deposition (MP4).

AUTHOR INFORMATION

Corresponding Author

* mboliveira@ua.pt, nmaraujo@fc.ul.pt, jmano@ua.pt

**Author Contributions**

The manuscript was written through contributions of all authors. All authors have given approval to the final version of the manuscript.

**Notes**

The authors declare no competing financial interest.

ACKNOWLEDGMENT

The authors acknowledge the financial support given by the Portuguese Foundation for Science and Technology (FCT) with the project "CellFi" (PTDC/BTM-ORG/3215/2020), and the European Research Council for the project "ATLAS" (ERC-2014-AdG-669858). This work was developed within the scope of the project CICECO-Aveiro Institute of Materials, UIDB/50011/2020 & UIDP/50011/2020 & LA/P/0006/2020, financed by national funds through the FCT/MCTES (PIDDAC).We also acknowledge financial support from the Portuguese

Foundation for Science and Technology (FCT) under Contracts no. PTDC/FIS-MAC/28146/2017 (LISBOA-01-0145-FEDER-028146), PTDC/FIS-MAC/5689/2020, UIDB/00618/2020, UIDP/00618/2020, CEECIND/00586/2017, SFRH/BD/143955/2019. A.F. Cunha acknowledges the FCT for the doctoral grant 2021.04817.BD. M.B. Oliveira acknowledges the contract 2021.03588.CEECIND.

For Table of Contents Only

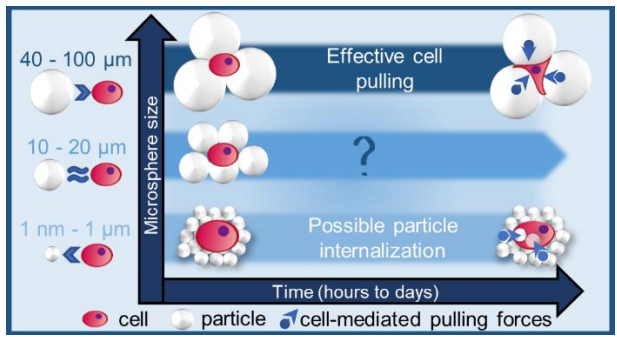